# METAMATERIALS FOR BALLISTIC ELECTRONS

D. Dragoman – Univ. of Bucharest, Physics Dept., P.O. Box MG-11, 077125 Bucharest, Romania

M. Dragoman[*] – National Institute for Research and Development in Microtechnology (IMT), P.O. Box 38-160, 023573 Bucharest, Romania

**Abstract**

The paper presents a metamaterial for ballistic electrons, which consists of a quantum barrier formed in a semiconductor with negative effective electron mass. This barrier is the analogue of a metamaterial for electromagnetic waves in media with negative electrical permittivity and magnetic permeability. Besides applications similar to those of optical metamaterials, a nanosized slab of a metamaterial for ballistic electrons, sandwiched between quantum wells of positive effective mass materials, reveals unexpected conduction properties, e.g. single or multiple room temperature negative differential conductance regions at very low voltages and with considerable peak-to-valley ratios, while the traversal time of ballistic electrons can be tuned to larger or smaller values than in the absence of the metamaterial slab. Thus, slow and fast electrons, analogous to slow and fast light, occur in metamaterials for ballistic electrons.

Author to whom the correspondence should be addressed; electronic mail mdragoman@yahoo.com, mircead@imt.ro

Metamaterials, or left-handed materials, commonly denote artificial structures excited by harmonic electromagnetic waves, which display both negative effective dielectric permittivity $\varepsilon$ and magnetic permeability $\mu$ parameters in a certain frequency bandwidth. In these structures the propagation vector ***k*** is opposite to the Poynting energy vector, contrary to the large category of right-handed electromagnetic media where both $\varepsilon$ and $\mu$ are positive and ***k*** has the same direction as the Poynting energy vector. For recent reviews on metamaterials see Refs. 1-2. Among the spectacular applications of left-handed materials we mention the superlens, which is a slab of metamaterial that focuses electromagnetic waves overcoming the diffraction limit, and the promising artificial magnetic materials in terahertz and optics range.

Ballistic electrons behave like matter waves, propagating without collisions in quantum structures over distances that can reach and even exceed 1µm at room temperature in some two-dimensional electron gases. Thus, typical micron- or millimetre-size devices for coherent electromagnetic waves such as lenses, waveguides, couplers, etc., can be imitated for ballistic electrons at a greatly reduced scale, i.e. at the nanoscale. For a review of ballistic electron devices and their properties see Ref. 3.

Recently, it was conjectured that metamaterials for ballistic electrons could exist in exotic structures such as twin boundaries of uniaxial semiconductors[4] or complementary media formed at the interface between two-dimensional graphitic lattices[5]. However, none of these implementations have any direct correspondence to the metamaterial for electromagnetic waves, characterized by the negative sign of both electrical permittivity and magnetic permeability. It is the main aim of this paper to establish such a correspondence based on the analogies reported in Ref. 6, and to explore some properties of the electronic metamaterials that have not been evidenced in the case of optical metamaterials.

For this purpose we consider a harmonic plane wave with angular frequency $\omega$ that propagates in the $z$ direction with field components $E_x$ and $H_y$. Then, the Maxwell equations have the simple form

$$\frac{d}{dz}\begin{pmatrix} E_x \\ H_y \end{pmatrix} = \begin{pmatrix} 0 & i\omega\mu \\ i\omega\varepsilon & 0 \end{pmatrix} \begin{pmatrix} E_x \\ H_y \end{pmatrix}. \qquad (1)$$

On the other hand, denoting by $\psi$ the wavefunction of a ballistic electron with kinetic energy $E$ and effective mass $m$, and by $\phi = d\psi/dz$ its spatial derivative, the quantum behaviour of the electrons in a region characterized by the potential energy $V$ can be described by the time-independent Schrödinger equation, expressed as

$$\frac{d}{dz}\begin{pmatrix} \psi \\ \phi \end{pmatrix} = \begin{pmatrix} 0 & m/\hbar \\ 2(E-V)/\hbar & 0 \end{pmatrix} \begin{pmatrix} \psi \\ \phi \end{pmatrix} \qquad (2)$$

The similarity between (1) and (2) suggests that the simultaneously negative sign of $\varepsilon$ and $\mu$ is analogous to the negative effective mass $m$ of ballistic electrons and the negative sign of $(E-V)$. Thus, the analogue of a left-handed material excited by electromagnetic waves is simply a barrier in a material with negative effective mass traversed by ballistic electrons. This analogy is also in agreement with the results reported in Ref. 6. As pointed out in Ref. 7, a negative effective mass of ballistic electrons can be found in common semiconductors, frequently used in quantum heterostructures, such as GaN, AlN, In$_{0.53}$Ga$_{0.47}$As, InAs, or InP with a certain crystallographic orientation. In principle, electronic metamaterials have similar properties to those of optical metamaterials. In particular, we expect that a metamaterial slab focuses the wavefunction of ballistic electrons beyond the diffraction limit and that at a metamaterial-right-hand material interface negative refraction of ballistic electrons occur (the last property was, in fact, used to find the exotic electronic metamaterials in Refs. 4 and 5).

However, in the following we would like to explore other characteristics of electronic transport through metamaterials easier to connect to electronic measurements (for example, the *I-V* characteristic) or to high-frequency or non-stationary operations (such as the traversal time). For this reason we consider throughout this paper a structure consisting of a slab of a metamaterial (an energetic barrier for electrons in a material with a negative mass) sandwiched between homogenous media, and assign to this multilayered structure, i.e. homogenous medium/metamaterial/homogenous medium, parameters labelled by the subscripts 1, 2 and 3, respectively. Medium 2 (the metamaterial) is encompassed between the planes $z = 0$ and $z = d$. Thus the wavefunction is given by

$$\psi = \begin{cases} A_1 \exp(ik_1 z) + B_1 \exp(-ik_1 z), & z < 0 \\ A_2 \exp(ik_2 z) + B_2 \exp(-ik_2 z), & 0 \leq z \leq d \\ A_3 \exp(ik_3 z), & z > d \end{cases} \qquad (3)$$

where $A_i$, $B_i$, $i = 1,2,3$, are the plane-wave amplitudes of counterpropagating wavefunction components in each layer and the wavenumbers $k_i$ are given by $k_{1,3} = [2m_{1,3}(E - V_{1,3})]^{1/2}/\hbar$, $k_2 = -[2m_2(E - V_2)]^{1/2}/\hbar$ (the negative sign in $k_2$ is needed, as in optics, to assure the same propagation direction of the wavefunction and of the corresponding quantum probability current density). Note that, since $m_2$ is negative, in the metamaterial slab $k_2$ is real (the wavefunction is propagating) for electron energies lower than the potential barrier $V_2$, the quantum wavefunction becoming evanescent (imaginary $k_2$) for higher electron energies; this behaviour is contrary to that encountered in materials with positive effective electron mass. The energetic barrier in metamaterials is analogous to an energetic well in metamaterials with positive effective electron mass.

The unknown wavefunction amplitudes in each layer are determined from the continuity conditions for the wavefunction and its derivative scaled by the effective mass at the frontiers $z = 0$ and $z = d$. The transmission coefficient, defined by

$$T(E) = k_3 \mid A_3 \mid^2 m_1 /(m_3 \mid A_1 \mid^2 k_1) \tag{4}$$

is then given by

$$\begin{aligned}T(E) = (4k_3m_1/m_3k_1)/[&\cos^2(k_2d)(1 + k_3m_1/m_3k_1)^2 \\ &+ \sin^2(k_2d)(k_3m_2/m_3k_2 + k_2m_1/m_2k_1)^2]\end{aligned} \tag{5}$$

For $m_1 = m_3 = 0.4m_0$, $m_2 = -0.02m_0$, where $m_0$ is the free electron mass, and $V_1 = V_3 = 0$, $V_2 = 0.5$ eV, the dependence of the transmission coefficient on the electron energy is displayed in Fig. 1(a) and 1(b) for various barrier widths. In Fig. 1(a) the solid line corresponds to $d = 5$ nm and the dotted line is assigned to $d = 15$ nm, whereas in Fig. 1b the solid line corresponds to $d = 30$ nm and the dotted line represents the case $d = 34$ nm. From these figures it follows that the transmission coefficient has significant values for electron energies comparable or lower than the barrier height $V_2$ in the metamaterial slab, the number of peaks in the transmission increasing with the barrier width; the transmission coefficient curve becomes steeper at the higher-energy side when the barrier width increases.

The corresponding *I-V* characteristics for the four transmission curves in Fig. 1, calculated with the Landauer formula at room temperature and for a Fermi energy level $E_F = 0$, are represented in Fig. 2. (In the simulations in Fig. 1 no applied bias was assumed.) We can see significant negative differential conductance regions in all cases. For $d = 5$ nm the current has a single negative differential conductance region with an impressive (practically infinite) peak-to-valley ratio. As the width of the barrier increases the number of peaks increases also, and multiple negative differential conductance regions are formed; the peak-to-valley ratio decreases in the multiple-peak case since the multiple resonances in transmission are crowded in the same energy interval, but still retains useful values. While a single negative differential conductance region is the key element of high frequency resonant tunnelling diode oscillators[8], the multiple negative differential conductance regions are essential for multi-valued logic elements in ultrafast computation schemes and for the most advanced logic gates based on them (see Ref. 9 and the references therein). Note that the metamaterial slab sandwiched between positive effective electron mass materials allows significant current flow only for low values of the applied voltage, having in this respect a unique *I-V* characteristic, difficult if not impossible to match by other similar structures. The metamaterial slab should then be used in low-power electronic circuits or as limiting device in common electronic circuits.

Further, we calculate the traversal time through the metamaterial slab according to

$$\tau = \int_0^d dz / v_g(z) \tag{6}$$

where $v_g(z) = J / |\psi(z)|^2$, $J = (\hbar i / 2m_2)[\psi(d\psi^* / dz) - \psi^*(d\psi / dz)]$, is the group velocity of ballistic electrons. After simple calculations we obtain

$$\tau = (m_3 / 2\hbar k_3)[(1 + \alpha^2)d + (1/2k_2)(1 - \alpha^2)\sin(2k_2 d)]. \tag{7}$$

where $\alpha = k_3 m_2 / (m_3 k_2)$. The traversal time for the metamaterial slab with $d = 5$ nm, as a function of the electron energy, has been represented with solid line in Fig. 3(a). With dotted and dashed lines in the same figure were displaced the traversal times for ballistic electrons through the same distance when the metamaterial slab is absent (i.e., when the electrons propagate in a homogeneous medium with $m_1 = m_2 = m_3 = 0.4m_0$, $V_1 = V_2 = V_3 = 0$) and when the metamaterial is unbounded (i.e., when no reflections at the metamaterial/right-handed material interfaces are considered), respectively. These times are equal to $\tau_{no\ slab} = dm_3/(\hbar k_3)$ and $\tau_{no\ refl} = dm_2/(\hbar k_2)$, respectively. It is interesting to note that, for almost the entire energy interval, $\tau < \tau_{no\ slab}$, implying that the metamaterial slab accelerates the ballistic electrons; this phenomenon is similar to the superluminal propagation of electromagnetic waves. This behaviour is maintained for larger barrier widths, as can be seen from Fig. 3(b) which displays the electron energy dependence of the same parameters for $d = 30$ nm. Moreover, in both cases $\tau > \tau_{no\ refl}$ for almost all energy interval. In both Figs. 3(a) and 3(b) one detects an energy value for which all three time parameters become equal: for $V_3 = 0$, as in the example considered in the paper, this is given by $E_{eq} = V_2 m_3/(m_3 - m_2)$, for which $\alpha = 1$. Above $E_{eq}$, $\tau > \tau_{no\ slab}$ and $\tau < \tau_{no\ refl}$, the presence of surrounding positive effective electron mass media speeding up ballistic electrons in thin metamaterial slabs; this behaviour, however, cannot be fully appreciated if $m_3 >> m_2$ (as in our example), since then $E_{eq} \cong V_2$. On the contrary, if $m_3 = |m_2| = 0.02 m_0$, $E_{eq} = V_2/2$ and the metamaterial slab can accelerate or delay the ballistic electrons with respect to the situation when it is absent. This behaviour is illustrated in Fig. 3(c), where the two regimes can be attained by ballistic electrons with energies lower or higher, respectively, than $E_{eq}$; the corresponding fast and slow electrons are analogous to fast and slow light, but with the important difference that the transition from one regime to another can be made continuously, in the same structure and with no need for additional fields, by simply adjusting the Fermi level. The possibility of tuning the traversal time of ballistic electrons that traverse a thin metamaterial slab by simply modifying the Fermi level is another unique characteristic of electronic metamaterials, with no equivalent (at least, to our knowledge) in electronics or in optics, for electromagnetic metamaterials. An even simpler way to tune the traversal time of ballistic electrons and, in particular, to tune the accelerating or delaying behaviour of the metamaterial slab with respect to the case when it is absent is to apply a voltage on the structure. The resulting voltage-dependent times $\tau$, $\tau_{no\ slab}$, and $\tau_{no\ refl}$ are represented in Fig. 3(d) with solid line, dotted line and dashed line, respectively, for $m_1 = m_3 = 0.4m_0$, $m_2 = -0.02m_0$, $V_1 = V_3 = 0$, $V_2 = 0.5$ eV, $d = 5$ nm, and for an electron energy of $E = 0.2$ eV. (In Figs. 3(a)-3(c) no applied voltage was assumed.) The tunable-delay ballistic electrons could find unexpected and unique applications in electronic or optoelectronic devices.

In conclusion we have shown that an energetic barrier for ballistic electrons in a material with a negative effective electron mass is the analogue of a metamaterial for electromagnetic radiation, and therefore should exhibit all the properties of optical metamaterials, such as perfect lensing of a slab, negative refraction, etc. Besides them, it is expected that specific electronic characteristics such as the *I-V* curve and the traversal time have unique properties for ballistic electrons propagating through metamaterials. These characteristics have been studied in this paper with astonishing results: The transmission coefficient has high values only for low-energy ballistic electrons, and hence the *I-V* curve is significant only for low biases, both characteristics exhibiting an increasing number of peaks with an increase in the thickness of the metamaterial slab. Both single- and multiple-peak *I-V* curves, with significant peak-to-valley ratio and with specific applications, can be obtained in low-power circuits. These characteristics are not encountered in common electronic structures that use positive effective electron mass materials. But the most astonishing property of a metamaterial slab is that it can accelerate or delay the traversal time of ballistic electrons with respect to the case in which it is absent. The traversal time tuning can be done by varying the Fermi level or, simpler, by applying a voltage on the right-hand material/metamaterial/right-hand material structure. The unique conduction and temporal characteristics of electronic metamaterial slabs could enhance their applications beyond the already astonishing world of mesoscopic devices, with particular emphasis on low-power electronics and electronic computing.

**FIGURE CAPTIONS**

**Fig. 1**  Electron energy dependence of the transmission coefficient through a metamaterial slab with a thickness of: (a) $d = 5$ nm (solid line), $d = 15$ nm (dotted line), (b) $d = 30$ nm (solid line), $d = 34$ nm (dotted line). See text for other parameters in the simulation.

**Fig. 2**  I-V curves for the corresponding situation in Fig. 1.

**Fig. 3**  Traversal time through a metamaterial slab (solid line), and for the same distance through a homogeneous right-hand surrounding medium (dotted line) and a homogeneous metamaterial (dashed line) for the structure in Fig. 1: (a) with $d = 5$ nm, (b) with $d = 30$ nm, (c) with $d = 5$ nm but equal modulus of positive and negative effective electron masses, (d) with $d = 5$ nm and applied voltage.

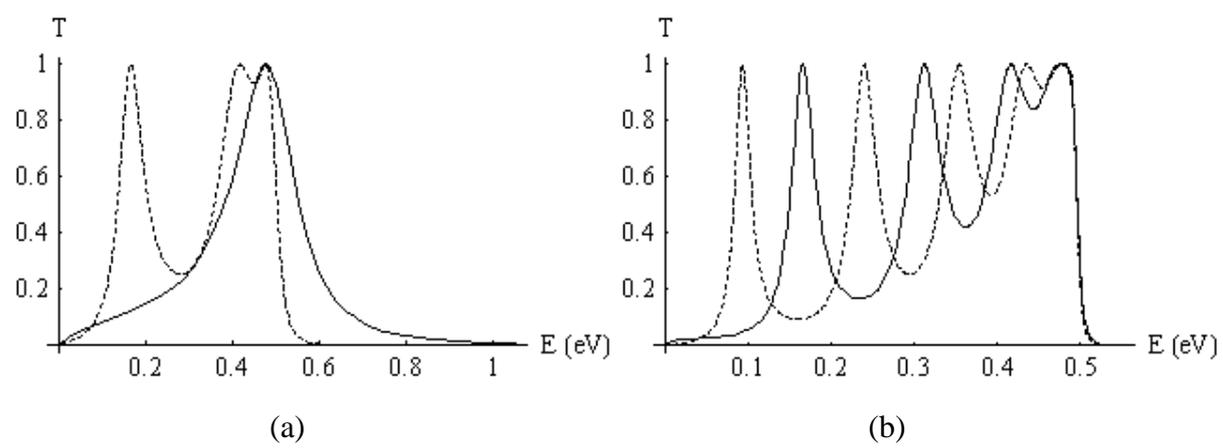

(a)                                           (b)

Fig.1

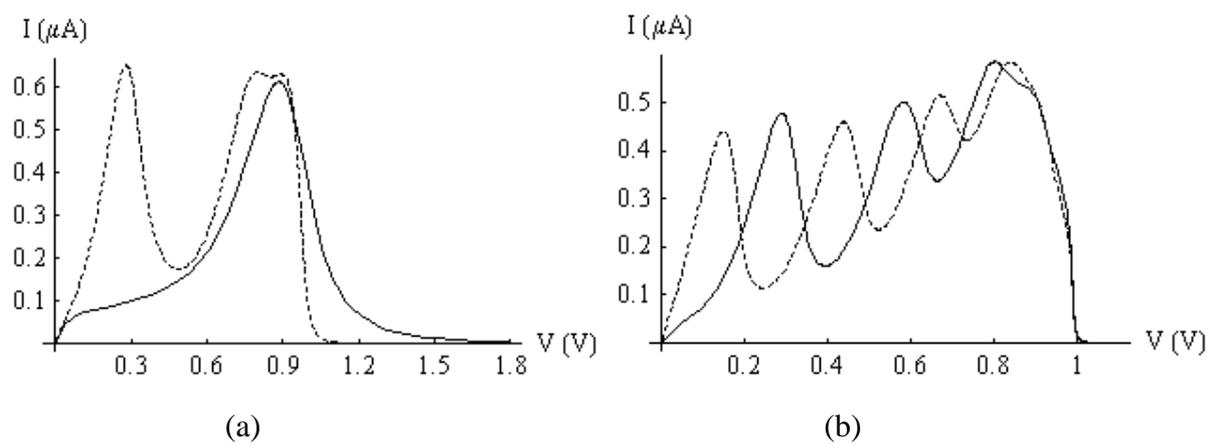

(a)                  (b)

Fig.2

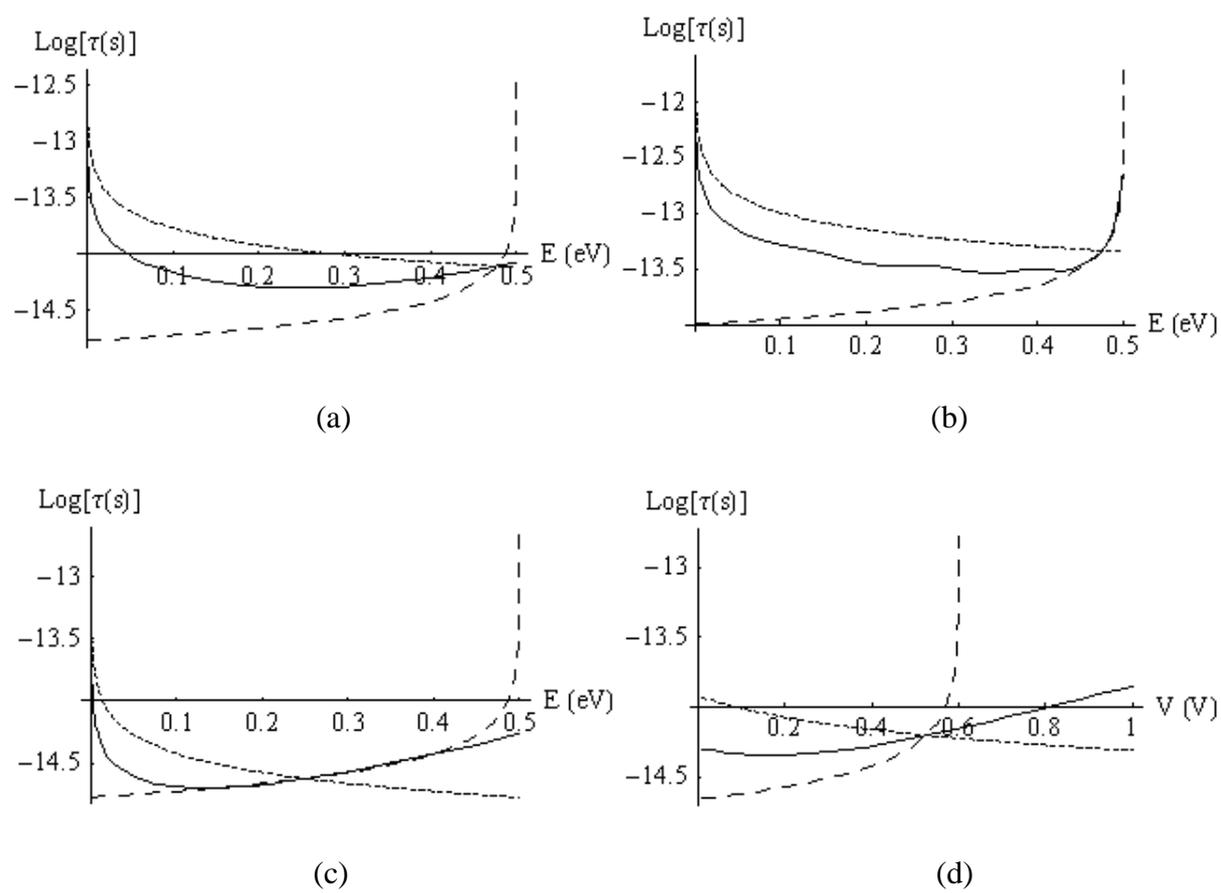

(a)　　　　　　　　　　　　　(b)

(c)　　　　　　　　　　　　　(d)

Fig.3